\documentclass{vgtc}                          

\graphicspath{{figures/}{pictures/}{images/}{./}} 

\usepackage{times}                     

\usepackage{tabu}                      
\usepackage{booktabs}                  
\usepackage{lipsum}                    
\usepackage{mwe}                       
\usepackage{cite}

\usepackage{mathptmx}                  
\usepackage{kotex}

\usepackage{booktabs}
\usepackage{threeparttable}
\usepackage{makecell}
\usepackage{rotating}   
\usepackage{tabularx}
\usepackage{siunitx}
\usepackage{graphicx}

\usepackage{verbatim}

\onlineid{1447}

\vgtccategory{Research}

\vgtcinsertpkg

\newcommand{\papername}{Point\&Spawn\xspace}
\title{\papername: Mid-Air Reference-Free Object Instantiation Using\\Gaze and Hand Gestures in Extended Reality}

\author{Jihyeon Lee\thanks{e-mail: jihyeon@kaist.ac.kr}\\ %
        \scriptsize KAIST %
\and Ken Pfeuffer\thanks{e-mail: ken@cs.au.dk}\\ %
     \scriptsize Aarhus University %
\and Jinwook Kim\thanks{e-mail: jinwook.kim31@kaist.ac.kr, Corresponding Author}\\ %
     \scriptsize KAIST %
\and Jeongmi Lee\thanks{e-mail: jeongmi@kaist.ac.kr, Corresponding Author}\\ %
      \scriptsize KAIST}

\teaser{
  \centering
  \includegraphics[width=0.98\linewidth]{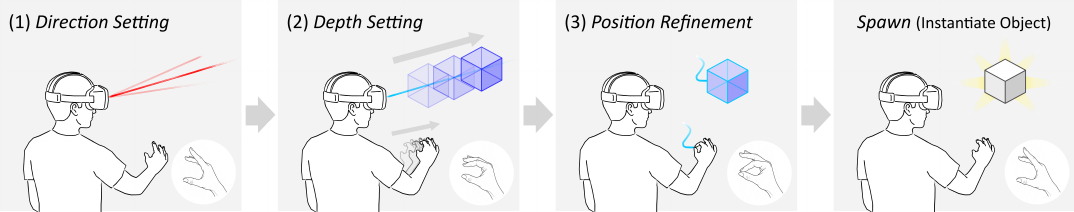}
  \caption{Illustration of the Point\&Spawn pipeline. \papername pipeline consists of three stages: (1) Direction Setting, (2) Depth Setting, and (3) Position Refinement, followed by object instantiation (spawn) upon pinch release. In this example, the user first sets the spawning direction with Gaze, then adjusts object depth using Relative Gain, refines the object position, and finally instantiates the object. The inset circles show the dominant hand gesture used at each stage: release for direction setting, semi-pinch for depth setting, pinch for position refinement, and release to confirm spawning.}
  \label{fig:teaser}
}

\abstract{
Mid-air object instantiation in XR requires users to specify a 3D position without spatial references, such as surfaces or existing objects. We present \papername, a staged pipeline for pre-instantiation position specification through Direction Setting, Depth Setting, and Position Refinement within a continuous gesture flow. We evaluated six controller-free techniques combining Gaze or Non-Dominant Hand (NDH) direction setting with Ray Intersection, Relative Gain, or Drag\&Hold depth setting in a user study (N=24) across Near and Far spawn depths. Relative Gain and Drag\&Hold yielded faster and more accurate spawning, lower workload, higher usability, and greater preference than Ray Intersection. The shoulder-referenced NDH ray improved speed and coarse accuracy, whereas the viewpoint-based Gaze ray reduced hand movement with comparable final accuracy. Farther spawn depth imposed greater temporal costs as well as Gaze and accuracy costs with Ray Intersection. These findings offer empirical guidance for designing direction and depth control in spawning in XR.
} 


\begin{document}


\firstsection{Introduction}
\maketitle
In extended reality (XR) environments, the instantiation and anchoring of digital objects at target locations constitute a fundamental and ubiquitous interaction~\cite{rau2025traversing, gonzalez2024guidelines}. Users frequently configure personalized workspaces by spawning and spatially arranging diverse digital elements, such as floating interfaces, annotations, and 3D artifacts, to support their ongoing tasks~\cite{cheng2025augmented, zhou2022depth, cools2025comparison}. Crucially, this workflow extends beyond the conventional selection and manipulation of pre-existing entities; it inherently encompasses the dynamic instantiation and precise spatial registration of new 3D objects and interfaces. Representative scenarios include mounting a virtual dashboard onto a physical wall, positioning a synthetic lighting model for interior design simulation, or displaying floating work instructions in mid-air near the user's active workspace while following a task~\cite{james2023evaluating, natephra2017integrating, shen2024application}.

When target locations are sufficiently proximal or provide a visual or physical reference, users can unambiguously specify the desired positions via direct touch or ray-casting (e.g., Gaze+Pinch, Hand Ray)~\cite{kim2023exploration, pfeuffer2017gaze, wagner2025study}. Accordingly, pointing, selection, and subsequent manipulation interaction techniques that target pre-existing objects by leveraging their colliders have been extensively studied, largely grounded in Fitts' law-based tasks~\cite{wagner2023fitts, mutasim2021pinch, kim2026align}.
In contrast, creating a new object presents a different challenge: it requires users to define, rather than select, its target location. The intended location may lie in empty mid-air, where no surfaces or existing objects are available as spatial references. In such cases, intuitively and accurately defining coordinates in 3D space, particularly along depth, remains difficult~\cite{bowman1997evaluation}. 
Consequently, most current systems initially instantiate objects at a default location (e.g., in front of the user), necessitating subsequent repositioning to the target coordinates via controller inputs or pinch gestures~\cite{rau2025traversing, hellmuth2021interaction}. 
This instantiate-then-reposition workflow separates object instantiation from spatial registration and can require additional interaction, particularly during distal placements. This motivates pre-instantiation position specification, in which users define and refine the intended position before the object is created.



Thus, we propose \papername, a staged interaction pipeline for mid-air object spawning using a single, continuous gesture flow in the absence of reference points. To mitigate the ambiguity of mid-air spatial targeting, we structured pre-instantiation position specification into three stages: \textit{Direction Setting, Depth Setting, and Position Refinement}. Specifically, following Guiard's bimanual interaction principles~\cite{guiard1987asymmetric}, \papername establishes the macro-level direction using either Gaze or the Non-Dominant Hand (NDH), while the Dominant Hand (DH) controls depth adjustment and final position refinement through semi-pinch and pinch gestures.

\noindent\textbf{Direction Setting:} We utilize either a viewpoint-based Gaze ray or a shoulder-referenced NDH ray to establish the initial direction. This design aimed to compare the trade-off between rapid, low-hand-involvement direction specification with Gaze and explicit, body-referenced directional control with NDH.

\noindent \textbf{Depth Setting:} To mitigate the Midas Touch problem~\cite{jacob1990you, penkar2012designing}, we employ a semi-pinch gesture with the DH to transition from direction setting to depth setting without immediately committing the position. We evaluate an established absolute ray-casting approach (Ray Intersection)~\cite{wyss2006isith, zhang2022conductor} against two relative depth-control methods (Relative Gain and Drag\&Hold) that map hand displacement to depth changes~\cite{baloup2019raycursor}.

\noindent \textbf{Position Refinement:} We use a pinch to initiate position refinement rather than immediately instantiate the object. While maintaining the pinch, users can continuously correct its position through hand displacement, and releasing the pinch confirms the position and instantiates the object. This stage enables users to compensate for errors from the prior stages and precisely anchor the object at the intended coordinates before committing the final position.

This design aims to streamline mid-air spatial spawning by integrating 3D point specification and refinement within a continuous pre-instantiation interaction. Through a controlled user study, we evaluate six interaction techniques combining two direction setting modalities and three depth setting methods across two spawn depths (Near vs. Far), examining their performance and user experience trade-offs. Our results show that the shoulder-referenced NDH ray supported faster spawning and more accurate coarse placement. By contrast, the viewpoint-based Gaze ray reduced hand movement, while final accuracy did not significantly differ between the two. Additionally, relative depth setting methods generally outperformed Ray Intersection across spawning time, accuracy, workload, usability, and preference. Farther spawn depth imposed more pronounced temporal costs with Gaze and accuracy costs for Ray Intersection. Based on these findings, we derive design guidelines for reference-free mid-air object spawning.

The contributions of our work are as follows: (1) We propose \papername, a staged interaction pipeline for mid-air object spawning that supports pre-instantiation position specification and refinement in the absence of spatial references; (2) We design six controller-free interaction techniques that combine two direction setting modalities with three depth setting methods within a shared semi-pinch-to-pinch gesture flow; and (3) Through a controlled user study, we characterize how direction setting modality, depth setting method, and spawn depth affect spawning performance and user experience, and derive design guidelines for reference-free mid-air object spawning.


\section{Related Work}
\subsection{Object Spawning in 3D XR Environment}
Unlike target selection or pointing, object spawning involves both targeting a spatial coordinate and executing a commit action to instantiate a persistent virtual object~\cite{keil2019preparing}. Thus, users rarely execute a single pointing motion during object spawning; rather, they perform a sequential workflow of spatial refinement and final confirmation~\cite{hellmuth2021interaction}. Therefore, spawning efficacy depends not only on the core 3D coordinate generation mechanism but also on the holistic workflow design, including the implementation of visual previews, coarse-to-fine refinement stages, reliable commit actions, and post-spawn adjustability~\cite{kyto2018pinpointing}.

In many XR systems, object placement is structured as a staged process rather than direct instantiation at the final target position. Objects are typically spawned within an accessible, near-field region and subsequently repositioned to the desired location through interactions, such as gizmo manipulation or grab interactions ~\cite{shapesxr_gizmo, rau2025traversing, cools2025comparison}. For example, HoloLens 2 adopts a staged placement workflow, in which a 3D object first appears in front of the user and is then moved and released to its intended location using grab or ray-based manipulation~\cite{ms_guides_place_holograms}. While this approach ensures easy manipulation in the initial stage and allows for incremental refinement, the repeated execution of these steps can increase interaction time and effort, particularly when placing multiple objects at distant locations~\cite{hincapieramos2014consumed}.

Another approach explores various geometric and procedural strategies to stabilize virtual object instantiation. In AR/MR contexts, systems frequently reference physical structures (e.g., walls or planes) to provide alignment, snapping, and spatial constraints, thereby improving the predictability of object spawn outcome~\cite{nuernberger2016snaptoreality}. Other approaches frame instantiation as a layout-generation problem, utilizing predefined rules and constraints to suggest optimal candidate positions and elevate overall instantiation quality~\cite{gal2014flare, bazargani2025integrating}. Additionally, to address the frequent viewpoint shifts inherent to handheld interfaces, researchers have enhanced instantiation stability by decoupling viewpoint stabilization from the active manipulation stage~\cite{lee2009freeze}. Collectively, these studies improved object instantiation and positioning by leveraging environmental constraints, automation, or temporally decoupled interaction. However, their benefits often depend on the availability of reference surfaces, reliable scene understanding, or interface conditions specific to handheld devices. As a result, they are less applicable to a reference-free mid-air environment, where users must determine and commit a target position in open 3D space without external spatial anchors.



\subsection{3D Point Specification for Mid-air Interaction}
Specifying a point or selecting and manipulating an object in 3D space is a fundamental component of interaction in XR~\cite{bowman1997evaluation, wagner2023fitts}. However, unlike 2D displays, 3D space makes it difficult to reliably resolve a user's intention to a single position~\cite{poupyrev1996go}. The challenge becomes more pronounced for distant targets, as a single directional cue can correspond to multiple depth candidates~\cite{wagner2024eye, wang2025headdepth}. The ambiguity is further amplified in reference-free mid-air environments without external anchors, such as walls or environmental features, often reducing interaction stability~\cite{bowman1997evaluation, shi2023exploration, wyss2006isith}.

Prior studies explored geometric approaches for 3D point specification by intersecting two rays or projecting a ray onto a plane or wall~\cite{wyss2006isith, zhang2022conductor, hellmuth2021interaction, wagner2024gaze, bashar2025depth3dsketch}. For example, Conductor specifies a 3D point as the intersection of a dominant-hand ray with a virtual plane controlled by the NDH, whereas Gaze+Controller combines gaze direction with a controller ray to support depth specification in 3D space. These approaches are intuitive, as they directly reflect the user's aiming behavior. However, with greater target distance and precision demands, even small tremors or posture changes can lead to large positional error~\cite{kopper2010human}.

To reduce alignment effort, indirect and relative mapping techniques treat depth not as a value to be aligned directly, but as a parameter that can be adjusted independently~\cite{shi2023exploration, wagner2024eye, lystbaek2022gaze, wagner2024gaze}. HeadDepth and RayCursor preserve the directional cue while assigning depth control to a separate axis, such as user movement or cursor-based mapping~\cite{wang2025headdepth, baloup2019raycursor}. Gain-based approaches have also been proposed to remap distant space into a nearby region or to amplify user movement~\cite{chae2018wall, poupyrev1996go, liu2025glance, ha2014wearhand, jung2017boosthand}. Although these techniques broaden the controllable range, supporting fine-grained refinement for manipulation under amplified movement remains a key challenge.

While various mechanisms for 3D point specification have been proposed, most have been examined in the context of selecting and manipulating pre-existing objects. In object spawning, by contrast, users must define and confirm a position in empty space, and the same techniques may behave differently when embedded in a spawning workflow. Accordingly, the tradeoff between absolute position specification and relative depth control remains underexplored in reference-free mid-air object spawning. In this work, we examine how these two design strategies affect spawning performance and user experience.



\section{Design of Object Spawning Technique}
To support controlled object spawning in reference-free XR environment, we designed the interaction as a sequential process consisting of three consecutive stages (Figure~\ref{fig:teaser}): (1) \textit{Direction Setting}, which establishes the spawning direction, (2) \textit{Depth Setting}, which determines the object position along that direction, and (3) \textit{Position Refinement}, which supports final local correction before spawning the object. During object spawning, three DH gestures (\textit{Pinch}, \textit{Semi-pinch}, and \textit{Release Pinch}) are used to control stage transitions~\cite{kim2025pinchcatcher}. 
Here, \textit{Pinch} indicates that the thumb and index finger are fully touching, \textit{Semi-pinch} corresponds to a gap of $1-5~cm$ between the fingertips, and \textit{Release Pinch} indicates a separation greater than $5~cm$. Each gesture corresponds to a distinct stage, thereby clearly separating them while maintaining overall interaction continuity. We particularly used \textit{Semi-pinch} as an intermediate state because it provides a clear transition zone between active control and release, allowing users to adjust depth without immediately committing or canceling the spawn. This allowed users to adjust depth while retaining the option to revise the target position or cancel the spawn before object instantiation.

\subsection{Stage 1: Direction Setting}
The direction setting stage aims to rapidly establish an initial spawning direction. We designed two approaches for this stage, using representative controller-free inputs: a viewpoint-based Gaze ray (Gaze) and a shoulder-referenced NDH ray (NDH), with their direction cues (Gaze cursor or NDH ray) shown in white. By comparing these settings, we evaluate the trade-off between immediate, low-effort, attention-driven directional control with Gaze~\cite{jacob1991use, chatterjee2015gaze+} and deliberate and explicit body-referenced directional control with NDH~\cite{lee2025facilitating, wobbrock2009user} during early-stage spawning. Upon entering the depth setting stage, the established spawning direction is locked to prevent unintended alterations from subsequent hand motions or gaze shifts, and the direction cue changes from white to blue.

\subsubsection{Gaze}
In the Gaze condition, the spawning direction is defined by a ray cast from the user's viewpoint, with a ring-shaped cursor providing continuous visual feedback. While this attention-driven approach enables rapid and intuitive direction specification~\cite{pfeuffer2017gaze, pfeuffer2024design}, gaze input is inherently susceptible to involuntary eye movements and minor fluctuations. Consequently, it could introduce directional instability, particularly when users target distant regions or unintentionally shift their attention~\cite{hou2024unveiling, kim2023exploration}.

\subsubsection{Non-Dominant Hand (NDH)}
In the NDH condition, a shoulder-referenced ray is formed from the shoulder joint through the center of the NDH, and the ray defines the spawning direction~\cite{turnerhand, mikkelsen2025dof}. The user is not required to form a specific hand shape and may maintain a comfortable, natural posture. Using the NDH in this way provides an explicit embodied reference for directional control while minimizing interference with the DH gestures used in later stages. Since the direction is fixed once depth setting begins, the user does not need to continuously maintain the arm posture throughout the remainder of the task.

\begin{figure}[tbp]
  \centering
  \includegraphics[width=\linewidth]{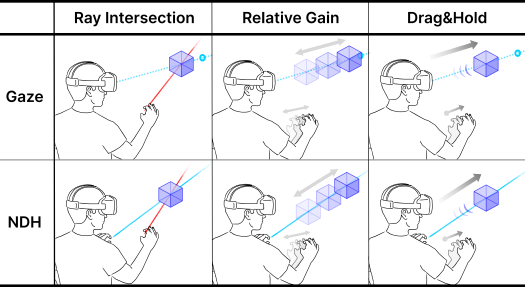}
  \caption{Illustration of the six object spawning techniques, combining two direction setting modalities (\textit{Gaze}, \textit{NDH}) and three depth setting methods (\textit{Ray Intersection} (Baseline), \textit{Relative Gain}, \textit{Drag\&Hold}).}
  \label{fig:technique}
\end{figure}

\subsection{Stage 2: Depth Setting}
Once the direction is defined, users adjust the object's depth along the locked ray (Figure~\ref{fig:technique}). This stage is initiated by a DH \textit{semi-pinch}, a posture where the thumb and index finger approach but do not fully close~\cite{zhu2023pinchlens, kim2025pinchcatcher}. We adopted this mechanism as it seamlessly integrates with the prevailing Gaze+Pinch interaction framework~\cite{pfeuffer2017gaze, pfeuffer2024design}. To enhance visual guidance, a semi-transparent blue preview cube indicates the provisional spawning location. Releasing the semi-pinch cancels the depth setting, reverting to the initial stage to re-specify the direction.

Here, we designed three depth setting methods (Figure~\ref{fig:technique}). \textit{Ray Intersection} specifies depth through absolute depth specification by intersecting two rays, whereas \textit{Relative Gain} and \textit{Drag\&Hold} use relative depth specification without requiring the user to directly align the DH ray with the locked direction ray. Between the two relative methods, \textit{Relative Gain} updates depth through incremental hand-motion-based adjustments, whereas \textit{Drag\&Hold} triggers continuous movement along the ray once the hand exceeds a displacement threshold. For the two relative methods, the preview cube initially appears at a fixed default point on the direction ray. The default position was set to 70 cm from the user, corresponding to an unreachable interaction distance~\cite{gonzalez2024guidelines}, to provide a common initial reference point. Subsequent depth changes are then computed relative to this point through DH motion.

\subsubsection{Ray Intersection (Baseline)}
Ray Intersection was used as the baseline because intersecting-ray approaches have been established in prior XR research, including Conductor and Gaze+Controller~\cite{zhang2022conductor, bashar2025depth3dsketch}. In this method, depth is determined by the intersection of the locked direction ray and a new ray cast from the DH. As in the NDH, we used a shoulder-referenced ray to reduce ray deviation during the transition to the refinement stage. Users specify depth by directly aiming the DH ray along the direction ray. The system computes the closest point between the two rays and validates the object only if the minimum distance is less than $0.2\,m$, a threshold matched to the side length of the cube. Its familiarity provides users with an intuitive means of specifying depth. However, it may become unstable or physically demanding when the target position is far away.

\subsubsection{Relative Gain}
Building on the RayCursor and Continuous Push technique for depth traversal and rate control principles~\cite{zhai1998user, lee2025facilitating, baloup2019raycursor}, Relative Gain maps DH motion to relative depth updates along the direction ray (Figure~\ref{fig:technique}). Instead of using the absolute hand position as the object position, the system interprets forward and backward DH motion as incremental input that updates the current depth. The magnitude of each update is modulated by movement speed, such that faster motion produces a larger depth change~\cite{zhai1998user}. Since the mapping is relative rather than absolute, users can cover a wider depth range with limited hand motion while still retaining fine-grained control over the object position.

\subsubsection{Drag\&Hold}
The Drag\&Hold was inspired by auto-scroll interactions in 2D web, in which an initial directional displacement triggers continuous movement while the input is maintained~\cite{aceituno2017design}. When the user moves the DH forward or backward by more than $3~cm$ (dead zone to prevent unintended activation) from the semi-pinch onset position, the preview cube begins to move continuously along the direction ray. Larger displacement beyond the threshold increases movement speed in real time, and returning the hand toward the initial position stops the movement. Unlike Relative Gain, which applies incremental updates from hand motion, this method uses displacement to initiate and regulate continuous movement. This enables efficient depth traversal with relatively small hand movements while preserving directional control through hand displacement.

\subsection{Stage 3: Position Refinement}
Once the depth is set, users perform a DH \textit{pinch}. Instead of immediately finalizing the spawn, the pinch initiates a refinement stage, visually indicated by a light-blue outline around the preview cube. While holding the pinch, users can adjust the object's position via a 1:1 mapping with their hand movement. Releasing the pinch commits the final location, instantiates the object, and concludes the spawn process. The integrated refinement stage is designed to maximize placement precision immediately before object creation~\cite{kyto2018pinpointing}.


\section{Evaluation}
We conducted a user study to evaluate the proposed object spawning techniques and address the following research questions.

\noindent\textbf{RQ1.}
How does the direction setting modality (\textit{Gaze} vs.\ \textit{NDH}) affect overall spawning performance and user experience across the object spawning pipeline?

\noindent\textbf{RQ2.}
How do absolute (\textit{Ray Intersection}) and relative (\textit{Relative Gain}, \textit{Drag\&Hold}) depth setting methods differ in performance and user experience in the spawning task?

\noindent\textbf{RQ3.}
How does spawn depth (\textit{Near} vs.\ \textit{Far}) affect spawning performance and effort across techniques, and which technique is most robust to these depth variations?


\begin{figure}[tbp]
  \centering
  \includegraphics[width= \linewidth]{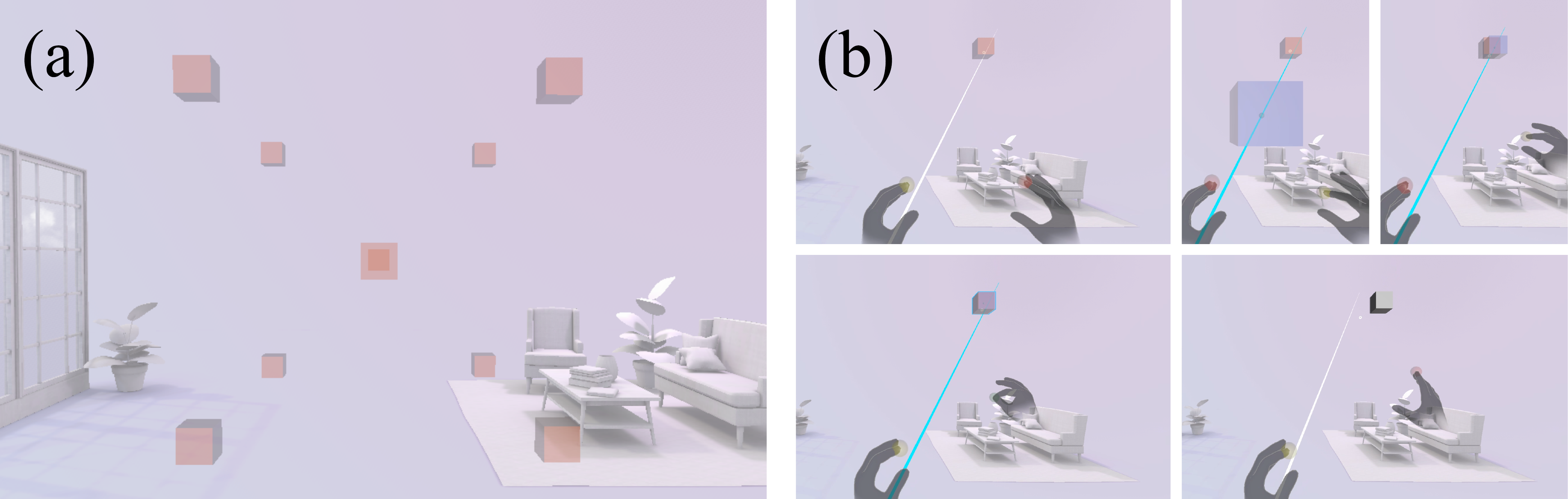}
  \caption{Experimental scenes used in the object spawning task. (a) Overview of all 10 candidate target positions. At each of the two spawn depths (\textit{Near}: $2.5~m$, \textit{Far}: $5.0~m$), targets could appear either at the center or at one of the four corners of a $2~m \times 2~m$ square. (b) Full interaction sequence for NDH+Drag\&Hold; read from left to right and then top to bottom.}
  \label{fig:exp_scene}
\end{figure}

\subsection{Study Design}
We evaluated the mid-air spawning performance of each technique in a \textit{single object spawning task}. The task was inspired by prior XR evaluation paradigms and adapted to our mid-air spawning scenario~\cite{wang2025headdepth, wagner2024eye}. The experiment employed a within-subject factorial design with 2 direction setting methods (NDH, Gaze) $\times$ 3 depth setting methods (Ray Intersection, Relative Gain, Drag\&Hold) $\times$ 2 spawn depths (Near, Far).

Participants used each interaction technique to spawn a virtual cube with a side length of $0.2\,m$ at a designated target position (Figure~\ref{fig:exp_scene} (a)). While users freely choose spawn positions in practical use cases, we used predefined locations to enable controlled measurements. Target positions comprised the center and four corners of a square region with side length $2\,m$, centered in front of the participant. The five candidate positions were arranged at two spawn depths: a near condition ($2.5\,m$), chosen to approximate the comfortable interaction distance of around $2\,m$ in XR, and a far condition ($5.0\,m$), representing a distant interaction range beyond hand reach~\cite{android2024spatialui, microsoft2019comfort}. We limited the study to these two representative depths to balance depth coverage and participant burden. This resulted in 10 candidate positions, each presented an equal number of times within a block. To clearly indicate the target position, we displayed a semi-transparent orange cube of the same size as the virtual cube to be spawned. The target cube served only as a visual ground-truth marker for controlled measurement and did not provide an environmental anchor.

Each trial began with gaze initialization by requiring participants to fixate on a cross mark presented in front of them for $200\,ms$. Once the mark disappeared, one of the 10 candidate positions was randomly selected, and the target position appeared with the constraint that the same position was not presented on consecutive trials. Participants were instructed to spawn an object at the target position as quickly and accurately as possible.

\subsection{Procedure}
Before the experiment, participants completed the consent forms and were briefed on the study purpose, task, and interaction techniques. They were then seated and equipped with a Meta Quest Pro headset, followed by a built-in eye-calibration procedure to ensure tracking accuracy. Before beginning the main experiment, a 20-minute training session was provided to allow participants to practice and familiarize themselves with the interaction techniques until they felt proficient.

The experiment consisted of six blocks, each corresponding to one interaction technique. The order of blocks was counterbalanced across participants using a balanced Latin square. Before each block, participants completed three practice trials. In the main experiment, they performed 20 trials at each spawn depth, yielding 40 trials per technique and 240 trials in total per participant.

After each block, participants completed questionnaires on task load, usability, and physical fatigue to evaluate the corresponding interaction technique. A short break was provided between blocks to reduce fatigue and minimize potential carry-over effects. After completing all blocks, they took part in a brief interview of approximately 5 minutes, which included a preference ranking of all techniques. The entire experiment lasted approximately 90 minutes.

\subsection{Apparatus}
We used a Meta Quest Pro (90 Hz, 106° horizontal × 96° vertical FoV) for the study. The virtual environment was developed in Unity 6 (6000.0.49f1) using the Oculus XR Plugin (v4.5.1), OpenXR Plugin (v1.14.3), and Meta XR Movement SDK (v83). Eye and hand input were provided by the headset’s built-in tracking sensors through the OVR camera rig. To improve the tracking input and reduce noise, we applied a 1\texteuro{} filter on gaze and hand movement~\cite{casiez20121, zhang2022conductor}. The parameters were set to $f_{c_{min}}$=0.9 and $\beta$=15 for gaze and $f_{c_{min}}$=0.9 and $\beta$=90 for hand movement, as in the previous Gaze-based interaction study~\cite{wagner2024eye}. The shoulder joint used to define the shoulder-referenced ray was obtained through OVRBody. The system ran on a desktop computer (AMD Ryzen 5 5600X, 32GB RAM, NVIDIA RTX 3060 Ti).

\subsection{Evaluation Metrics} 


\subsubsection{Objective Measures}

\noindent\textbf{Direction Setting Time}
refers to the cumulative duration from the trial onset until depth setting was initiated. This metric was computed by summing all intervals during which the DH remained in the release state before depth setting began.

\noindent\textbf{Depth Setting Time}
refers to the cumulative duration from the initiation of depth setting until the object depth was confirmed, thereby completing the coarse spawn. 
This metric was computed by summing all intervals during which the DH remained in the semi-pinch state during depth setting.

\noindent\textbf{Refinement Time}
refers to the duration during which the participant maintained the pinch from the completion of the coarse spawn until the object position was finalized and the trial ended.

\noindent\textbf{Coarse Spawn Time}
refers to the total duration from trial onset to the completion of the coarse spawn. This metric captures the overall time required for direction and depth setting up to the coarse spawn.

\noindent\textbf{Total Spawn Time} 
refers to the total duration from trial onset to the end of the trial, when the final object position was confirmed. This metric captures the overall time required to complete the entire spawning task.

\noindent\textbf{Coarse Offset}
refers to the Euclidean distance between the center of the coarse-spawned cube and the target position immediately after the coarse spawn was completed. This metric captures spawning accuracy before refinement.

\noindent\textbf{Final Offset}
refers to the Euclidean distance between the final position of the spawned cube and the target position at the end of the trial. This metric captures final spawning accuracy.

\noindent\textbf{Hand Movement} 
refers to the cumulative movement distance of both the dominant and NDHs over the entire trial. This metric captures the amount of physical motion required by each technique.

\subsubsection{Subjective Measures}
Subjective ratings were collected for each interaction technique, defined as the combination of direction setting modality and depth setting method, and were not collected separately by spawn depth.

\noindent\textbf{NASA Task Load Index (NASA-TLX)}
refers to the subjective workload~\cite{hart2006nasa}. Participants rated six dimensions (mental demand, physical demand, temporal demand, performance, effort, and frustration) on a 100-point scale with 5-point increments, and the ratings were averaged into a total workload score.

\noindent\textbf{System Usability Scale (SUS)}
refers to the usability~\cite{brooke2013sus}. The SUS consists of 10 items, and responses on a 5-point Likert scale were converted into a standard score ranging from 0 to 100.

\noindent\textbf{Physical Fatigue}
were measured using the Borg-CR10 scale~\cite{borg1982psychophysical}. Participants rated hand and eye fatigue separately on a scale from 0 to 10.

\noindent\textbf{Preference Ranking}
refers to the preference across the six techniques, collected after all tasks had been completed. Participants ranked the techniques from 1 (most preferred) to 6 (least preferred).

\subsection{Statistical Analysis}
After averaging trials per participant for each condition, normally distributed metrics (direction time, refinement time, and subjective measures other than preference ranking) were evaluated via repeated-measures (RM) ANOVA, with Greenhouse-Geisser corrections for sphericity violations. Non-normal data were analyzed using Aligned Rank Transform (ART) ANOVA~\cite{wobbrock2011aligned}. We used participant-specific error terms for all within-subject effects. For significant effects, we conducted Bonferroni-corrected post-hoc comparisons: paired \textit{t}-tests for RM ANOVA and estimated marginal means (\textit{emmeans}) on ART linear models for ART main effects. Significant ART interactions were followed by the ART-C procedure. We report partial eta-squared ($\eta_p^2$) and Cohen’s $d_z$ for effect sizes. Descriptive statistics reflect original scales, though ART-based partial eta-squared values rely on aligned ranks. Preference rankings were evaluated separately via a Friedman test with Nemenyi post-hoc comparisons. All analyses were conducted in R (v.4.4.1).


\subsection{Participants}
A total of 24 participants (12 male, 12 female; Mean age 25.17, $SD$ = 4.36) participated in the study and received \$20 compensation. Among them, 19 individuals had prior experience using VR more than five times, and nine had prior experience using Gaze+Pinch and pinch interactions more than five times. Regarding eyesight correction, one participant wore glasses, 15 reported contact lens use or prior eye surgery, and eight reported none. All participants were right-handed. The study protocols and methods were approved by the Institutional Review Board (IRB), and all participants provided written informed consent.


\begin{figure*}[!t]
  \centering
  \includegraphics[width=\linewidth]{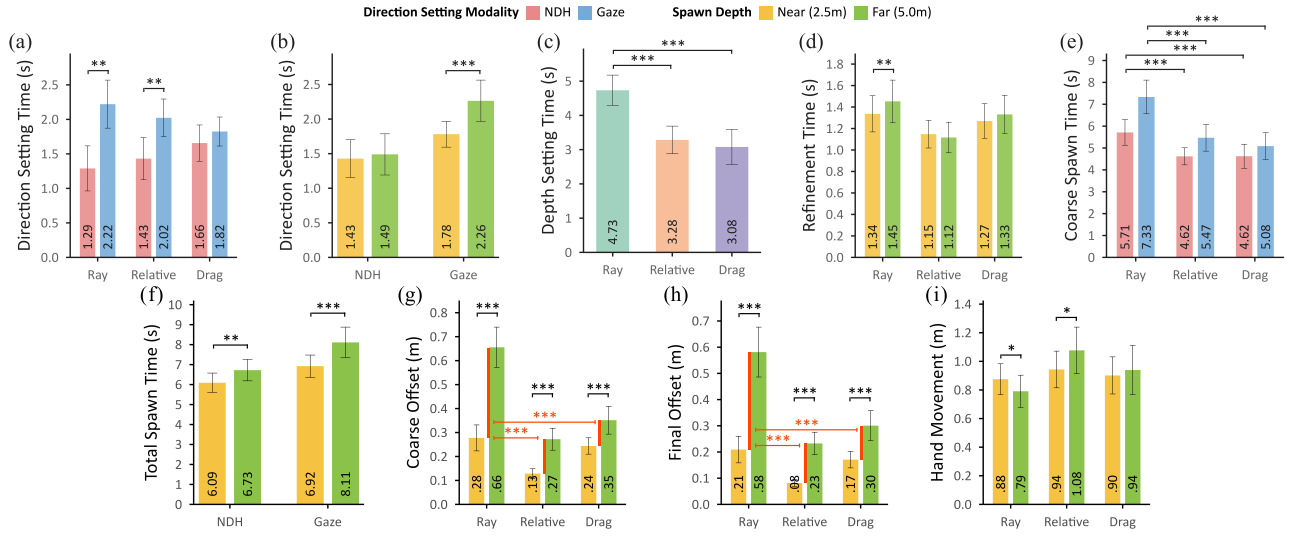}
  \caption{Plots for the objective measures with significant main or interaction effects: direction setting time (a) by direction setting modality and depth setting method and (b) by direction setting modality and spawn depth; (c) depth setting time by depth setting method; (d) refinement time by depth setting method and spawn depth; (e) coarse spawn time and (f) total spawn time by direction setting modality and spawn depth; (g) coarse offset and (h) final offset by depth setting method and spawn depth; (i) hand movement by depth setting method and spawn depth. Statistical significance is denoted by * for $p<.05$, ** for $p<.01$, and *** for $p<.001$. Error bars indicate 95\% confidence intervals.}
  \label{fig:graph_figure}
  \vspace{2pt}
  \raggedright\scriptsize
  \textit{Note.} Depth setting method labels are abbreviated as Ray (Ray Intersection), Relative (Relative Gain), and Drag (Drag\&Hold).
\end{figure*}

\section{Result}
Among the total of 5760 trials collected, 38 trials (0.66\%) were excluded due to abnormally short coarse spawn times ($<$ 1.0 second), which likely reflected accidental pinch inputs rather than intentional spawn behavior. Unless otherwise noted, all values are reported as mean $\pm$ standard deviation (SD).
Table~\ref{tab:statistics_table} summarizes the omnibus results for the objective measures. 

\begin{table*}[t]
\centering
\scriptsize
\setlength{\tabcolsep}{0.3pt}
\renewcommand{\arraystretch}{1.08}

\caption{Summary of omnibus tests for the objective measures. Degrees of freedom are shown below the corresponding $F$ values.}
\label{tab:statistics_table}

\newcommand{\Fdf}[2]{%
  \begin{tabular}[c]{@{}c@{}}\ensuremath{#1}\\[-2.5pt]{\tiny\ensuremath{(#2)}}\end{tabular}}
\newcommand{\FdfLong}[2]{\scalebox{.88}[1]{\Fdf{#1}{#2}}}
\newcommand{\pLT}{\resizebox{6mm}{!}{\ensuremath{\mathord{<}.001}}}

\begin{tabular}{@{}>{\raggedright\arraybackslash}m{20mm}
  *{5}{>{\centering\arraybackslash}m{12.5mm}
        >{\centering\arraybackslash}m{6.5mm}
        >{\centering\arraybackslash}m{5.5mm}
        @{\hspace{0.4mm}}}
        >{\centering\arraybackslash}m{12.5mm}
        >{\centering\arraybackslash}m{6.5mm}
        >{\centering\arraybackslash}m{5.5mm}@{}}
\toprule
& \multicolumn{3}{c}{{\scriptsize\bfseries Direction Modality}}
& \multicolumn{3}{c}{{\scriptsize\bfseries Depth Method}}
& \multicolumn{3}{c}{{\scriptsize\bfseries Spawn Depth}}
& \multicolumn{3}{c}{{\scriptsize\bfseries Modality$\times$Method}}
& \multicolumn{3}{c}{{\scriptsize\bfseries Modality$\times$Depth}}
& \multicolumn{3}{c}{{\scriptsize\bfseries Method$\times$Depth}} \\
\cmidrule(lr){2-4}\cmidrule(lr){5-7}\cmidrule(lr){8-10}
\cmidrule(lr){11-13}\cmidrule(lr){14-16}\cmidrule(lr){17-19}
& $F$ & $p$ & $\eta_p^2$
& $F$ & $p$ & $\eta_p^2$
& $F$ & $p$ & $\eta_p^2$
& $F$ & $p$ & $\eta_p^2$
& $F$ & $p$ & $\eta_p^2$
& $F$ & $p$ & $\eta_p^2$ \\
\midrule

\resizebox{19.5mm}{!}{{\scriptsize\bfseries Direction Time}}
& \Fdf{10.48}{1,23} & .004 & .31
& \FdfLong{0.08}{1.57,36.14} & n.s. & .00
& \Fdf{53.14}{1,23} & \pLT & .70
& \FdfLong{10.91}{1.62,37.20} & \pLT & .32
& \Fdf{15.67}{1,23} & \pLT & .41
& \FdfLong{0.35}{1.94,44.59} & n.s. & .02 \\
\addlinespace[1pt]

{\scriptsize\bfseries Depth Time}
& \Fdf{9.07}{1,23} & .006 & .28
& \Fdf{46.61}{2,46} & \pLT & .67
& \Fdf{70.64}{1,23} & \pLT & .75
& \Fdf{0.44}{2,46} & n.s. & .02
& \Fdf{0.30}{1,23} & n.s. & .01
& \Fdf{2.46}{2,46} & n.s. & .10 \\
\addlinespace[1pt]

\resizebox{19.5mm}{!}{{\scriptsize\bfseries Refinement Time}}
& \Fdf{9.40}{1,23} & .006 & .29
& \FdfLong{19.27}{1.91,43.85} & \pLT & .46
& \Fdf{4.61}{1,23} & .043 & .17
& \FdfLong{0.12}{1.65,37.84} & n.s. & .01
& \Fdf{3.94}{1,23} & n.s. & .15
& \FdfLong{5.46}{1.89,43.47} & .009 & .19 \\
\addlinespace[1pt]

{\scriptsize\bfseries Coarse Time}
& \Fdf{15.59}{1,23} & \pLT & .40
& \Fdf{41.55}{2,46} & \pLT & .64
& \Fdf{88.83}{1,23} & \pLT & .79
& \Fdf{5.32}{2,46} & .008 & .19
& \Fdf{7.39}{1,23} & .012 & .24
& \Fdf{0.69}{2,46} & n.s. & .03 \\
\addlinespace[1pt]

{\scriptsize\bfseries Total Time}
& \Fdf{16.26}{1,23} & \pLT & .41
& \Fdf{48.88}{2,46} & \pLT & .68
& \FdfLong{100.80}{1,23} & \pLT & .81
& \Fdf{2.92}{2,46} & n.s. & .11
& \Fdf{7.88}{1,23} & .010 & .26
& \Fdf{0.37}{2,46} & n.s. & .02 \\
\addlinespace[1pt]

{\scriptsize\bfseries Coarse Offset}
& \Fdf{4.52}{1,23} & .045 & .16
& \FdfLong{119.63}{2,46} & \pLT & .84
& \FdfLong{318.89}{1,23} & \pLT & .93
& \Fdf{0.91}{2,46} & n.s. & .04
& \Fdf{0.37}{1,23} & n.s. & .02
& \Fdf{49.98}{2,46} & \pLT & .68 \\
\addlinespace[1pt]

{\scriptsize\bfseries Final Offset}
& \Fdf{1.26}{1,23} & n.s. & .05
& \Fdf{98.15}{2,46} & \pLT & .81
& \FdfLong{417.66}{1,23} & \pLT & .95
& \Fdf{0.55}{2,46} & n.s. & .02
& \Fdf{0.04}{1,23} & n.s. & .00
& \Fdf{45.32}{2,46} & \pLT & .66 \\
\addlinespace[1pt]

\resizebox{19.5mm}{!}{{\scriptsize\bfseries Hand Movement}}
& \Fdf{14.50}{1,23} & \pLT & .39
& \Fdf{6.31}{2,46} & .004 & .22
& \Fdf{0.72}{1,23} & n.s. & .03
& \Fdf{1.08}{2,46} & n.s. & .05
& \Fdf{9.41}{1,23} & .006 & .29
& \Fdf{15.51}{2,46} & \pLT & .40 \\

\bottomrule
\end{tabular}

\vspace{2pt}
\raggedright\scriptsize
\textit{Note.} n.s. = not significant ($p\ge.05$). All three-way interactions were non-significant. Labels are shortened for brevity (e.g., ``setting'', ``spawn'' are omitted).
\end{table*}

\subsection{Spawning Efficiency (Figure~\ref{fig:graph_figure} (a-f))}
\subsubsection{Direction Setting Time}
The analysis revealed a significant main effect of direction setting modality, indicating that NDH was faster than Gaze (NDH: $1.46 \pm .73$; Gaze: $2.02 \pm .78$).
There was also a significant main effect of spawn depth, with longer direction setting times for far than near conditions (Near: $1.60 \pm .66$; Far: $1.88 \pm .91$). 
The main effect of the depth setting method was not significant ($p=.886$).

We found a significant interaction between direction setting modality and depth setting method. 
Modality comparisons within each method showed that Gaze was slower than NDH for Ray Intersection ($t(23)=3.87$, $p=.001$, $d_z=.79$) and Relative Gain ($t(23)=2.86$, $p=.009$, $d_z=.58$), whereas the modality difference was not significant for Drag\&Hold ($p=.209$).
Another significant interaction was found between direction setting modality and spawn depth. 
Within Gaze, direction setting took longer for far than near conditions ($t(23)=5.91$, $p<.001$, $d_z=1.21$), whereas the depth difference within NDH was not significant ($p=.168$).
There were no significant effects for the remaining interactions ($ps \ge .479$).

\subsubsection{Depth Setting Time}
The analysis revealed a significant main effect of direction setting modality, indicating that NDH was faster than Gaze (NDH: $3.50 \pm 1.36$; Gaze: $3.90 \pm 1.54$).
The main effect of the depth setting method was significant. Ray Intersection was slower than Drag\&Hold ($t(46)=9.12$, $p<.001$, $d_z=1.51$) and Relative Gain ($t(46)=7.30$, $p<.001$, $d_z=2.08$), whereas Drag\&Hold and Relative Gain did not differ significantly ($p=.222$; Ray Intersection: $4.73 \pm 1.32$; Relative Gain: $3.28 \pm 1.08$; Drag\&Hold: $3.08 \pm 1.39$).
Spawn depth showed a significant main effect, with longer depth setting times for far than near conditions (Near: $3.41 \pm 1.35$; Far: $3.99 \pm 1.52$).
There were no significant interactions ($ps \ge .097$).

\subsubsection{Refinement Time}
The analysis revealed a significant main effect of direction setting modality, indicating that NDH was faster than Gaze (NDH: $1.21 \pm .44$; Gaze: $1.34 \pm .44$).
Depth setting method showed a significant main effect. 
Post-hoc tests indicated that Relative Gain was faster than Drag\&Hold ($t(23)=4.08$, $p=.001$, $d_z=.83$) and Ray Intersection ($t(23)=5.55$, $p<.001$, $d_z=1.13$), whereas the difference between Drag\&Hold and Ray Intersection was not significant ($p=.080$) (Ray Intersection: $1.39 \pm .47$; Relative Gain: $1.13 \pm .36$; Drag\&Hold: $1.30 \pm .45$).
Spawn depth showed a main effect, with longer refinement times for far than near conditions (Near: $1.25 \pm .40$; Far: $1.30 \pm .48$).

We found a significant interaction between depth setting method and spawn depth. Depth comparisons within each method further showed that only Ray Intersection took longer for far than near conditions ($t(23)=3.02$, $p=.006$, $d_z=.62$), whereas the depth difference was not significant for Drag\&Hold ($p=.109$) or Relative Gain ($p=.255$).
Other interactions were not significant ($ps \ge .059$).

\subsubsection{Coarse Spawn Time}
The analysis revealed a significant main effect of direction setting modality, indicating that NDH was faster than Gaze (NDH: $4.98 \pm 1.38$; Gaze: $5.96 \pm 2.01$).
Depth setting method showed a significant main effect. Bonferroni-corrected post-hoc tests indicated that Ray Intersection was slower than Drag\&Hold ($t(46)=8.62$, $p<.001$, $d_z=1.47$) and Relative Gain ($t(46)=6.88$, $p<.001$, $d_z=1.82$), whereas the difference between Drag\&Hold and Relative Gain was not significant ($p=.268$) (Ray Intersection: $6.52 \pm 1.91$; Relative Gain: $5.04 \pm 1.39$; Drag\&Hold: $4.85 \pm 1.54$).
Spawn depth showed a significant main effect, with longer coarse spawn times for far than near conditions (Near: $5.05 \pm 1.54$; Far: $5.90 \pm 1.91$).

We found a significant interaction between direction setting modality and depth setting method. 
Within both modalities, Ray Intersection was slower than Drag\&Hold ($t$s$(115)\ge4.48$,\ $p$s$<.001$,\ $d_z$s$\ge.74$) and slower than Relative Gain ($t$s$(115)\ge4.12$,\ $p$s$<.001$,\ $d_z$s$\ge1.22$), whereas Drag\&Hold and Relative Gain did not differ ($p$s$\ge.394$).
We also found a significant interaction between direction setting modality and spawn depth. Coarse spawn time was longer for far than near conditions in both Gaze ($t(69)=3.93$,\ $p<.001$,\ $d_z=1.45$) and NDH ($t(69)=3.25$,\ $p=.002$,\ $d_z=1.74$), with a larger increase in Gaze.
No other interactions were significant ($ps \ge .508$).

\subsubsection{Total Spawn Time}
The analysis revealed a significant main effect of direction setting modality, indicating that NDH was faster than Gaze (NDH: $6.41 \pm 1.56$; Gaze: $7.52 \pm 2.21$).
Depth setting method showed a significant main effect. Ray Intersection was slower than Drag\&Hold ($t(46)=8.87$, $p<.001$, $d_z=1.49$) and Relative Gain ($t(46)=8.22$, $p<.001$, $d_z=2.06$), whereas Drag\&Hold and Relative Gain did not differ ($p=1$) (Ray Intersection: $8.13 \pm 2.12$; Relative Gain: $6.38 \pm 1.56$; Drag\&Hold: $6.38 \pm 1.72$).
Spawn depth showed a significant main effect, with longer total spawn times for far than near conditions (Near: $6.51 \pm 1.72$; Far: $7.42 \pm 2.13$).

We found a significant interaction between direction setting modality and spawn depth. Total spawn time was longer for far than near conditions in both Gaze ($t(69)=4.04,\ p<.001,\ d_z=1.53$) and NDH ($t(69)=2.97,\ p=.004,\ d_z=1.53$), with a larger increase in Gaze. No other interactions were significant ($ps \ge .064$).

\subsection{Spawning Accuracy (Figure~\ref{fig:graph_figure} (g-i))}
As defined in our metrics, the offset was computed as the Euclidean distance between the target position and the coarse or final object spawned position. Lower values indicate better accuracy.

\subsubsection{Coarse Offset}
The analysis revealed a significant main effect of direction setting modality, indicating that the coarse offset was smaller with NDH than with Gaze (NDH: $.305 \pm .203$; Gaze: $.338 \pm .238$).
Depth setting method showed a significant main effect. Ray Intersection yielded a larger offset than Drag\&Hold ($t(46)=8.42$, $p<.001$, $d_z=1.54$) and Relative Gain ($t(46)=15.45$, $p<.001$, $d_z=1.84$), and Drag\&Hold yielded a larger offset than Relative Gain ($t(46)=7.03$, $p<.001$, $d_z=1.06$) (Ray Intersection: $.467 \pm .269$; Relative Gain: $.200 \pm .141$; Drag\&Hold: $.298 \pm .138$).
Spawn depth showed a significant main effect, with larger coarse offsets for far than near conditions (Near: $.217 \pm .122$; Far: $.426 \pm .248$).

We found a significant interaction between depth setting method and spawn depth.
Depth comparisons within each method showed that the offset was larger for far than near conditions for Ray Intersection ($t(115)=11.40$,\ $p<.001$,\ $d_z=3.15$), Relative Gain ($t(115)=8.79$,\ $p<.001$,\ $d_z=1.22$), and Drag\&Hold ($t(115)=5.21$, $p<.001$, $d_z=1.09$). 
Since the interaction pattern was similar across methods, we further analyzed the magnitude of difference between the far and near conditions within each method ($\Delta_\text{far-near}$). Paired Wilcoxon signed-rank tests with Bonferroni correction showed that Ray Intersection produced a larger difference than the other two methods ($ps < .001$), whereas Relative Gain and Drag\&Hold did not differ significantly ($p = 1$).
There were no other significant interactions ($ps \ge .368$).

\subsubsection{Final Offset}
Depth setting method showed a significant main effect. Ray Intersection yielded a larger offset than Drag\&Hold ($t(46)=8.00,\ p<.001,\ d_z=1.42$) and Relative Gain ($t(46)=13.96,\ p<.001,\ d_z=1.53$), and Drag\&Hold yielded a larger offset than Relative Gain ($t(46)=5.96,\ p<.001,\ d_z=.82$) (Ray Intersection: $.395 \pm .276$; Relative Gain: $.157 \pm .139$; Drag\&Hold: $.236 \pm .140$).
Spawn depth showed a significant main effect, with larger final offsets for far than near conditions (Near: $.154 \pm .111$; Far: $.372 \pm .245$).
The main effect of direction setting modality was not significant ($p=.274$).

We found a significant interaction between depth setting method and spawn depth. 
Depth comparisons within each method showed that the offset was larger for far than near conditions for Ray Intersection ($t(115)=12.30$, $p<.001$, $d_z=2.68$), Relative Gain ($t(115)=10.60$, $p<.001$, $d_z=1.39$), and Drag\&Hold ($t(115)=7.28$, $p<.001$, $d_z=1.39$). 
Regarding  $\Delta_\text{far-near}$ within each method, Ray Intersection produced a larger difference than the other two methods ($ps < .001$), whereas Relative Gain and Drag\&Hold did not differ significantly ($p = 1$).
No other interactions were significant ($ps \ge .579$).

\subsection{Hand Movement (Figure~\ref{fig:graph_figure} (i))}
The analysis revealed a significant main effect of direction setting modality. Hand movement was larger for NDH than Gaze (Gaze: $.84m \pm .38$; NDH: $1.00m \pm .40$).
Depth setting method showed a significant main effect. Relative Gain resulted in larger hand movement than Ray Intersection ($t(46)=3.44$, $p=.004$, $d_z=.69$), whereas its difference from Drag\&Hold was not significant ($p = .051$). Drag\&Hold and Ray Intersection also did not differ ($p=1$) (Ray Intersection: $.83m \pm .30$; Relative Gain: $1.01m \pm .40$; Drag\&Hold: $.92m \pm .45$).
Spawn depth did not show a significant main effect ($p=.406$).

We found a significant interaction between direction setting modality and spawn depth. Hand movement was larger for NDH than Gaze within both far ($t(69)=3.08,\ p=.003,\ d_z=.52$) and near conditions ($t(69)=5.70,\ p<.001,\ d_z=.91$), with a larger modality difference in the near condition.
Another significant interaction was found between depth setting method and spawn depth. Depth comparisons within each method showed that hand movement was larger for near than far conditions for Ray Intersection ($t(115)=2.08$, $p=.040$, $d_z=.91$), whereas Relative Gain showed larger hand movement for far than near conditions ($t(115)=2.15,\ p=.034,\ d_z=.82$). The depth effect was not significant for Drag\&Hold ($p=.948$).
There were no other significant interactions ($ps \ge .091$).

\begin{figure*}[tbp]
  \centering
  \includegraphics[width=\linewidth]{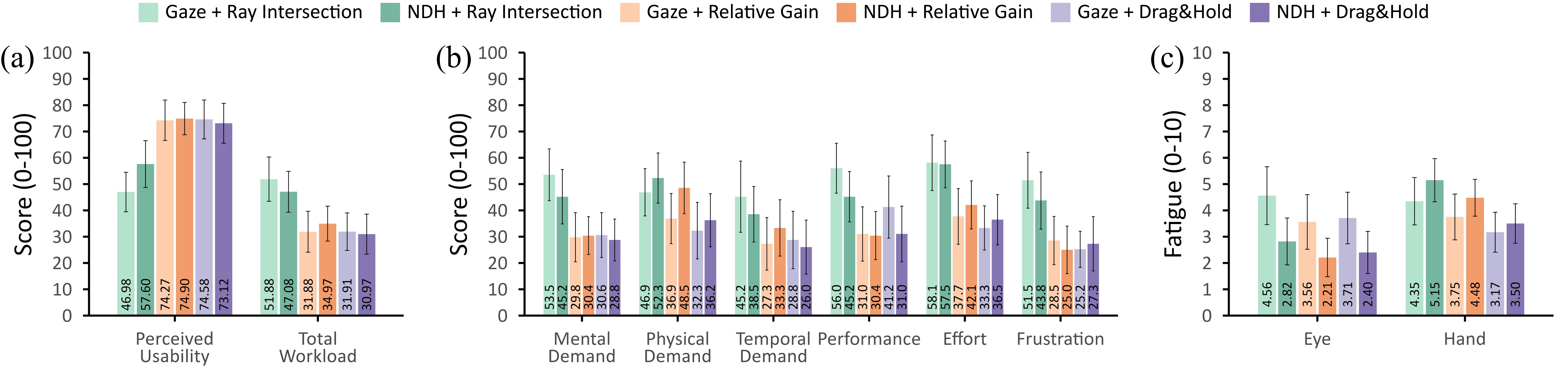}
  \caption{(a) Average perceived usability (SUS) and total task workload (NASA-TLX) across the six object spawning techniques (normalized to a 0–100 scale). (b) Average scores for the six NASA-TLX dimensions. (c) Average eye and hand fatigue scores (Borg-CR10 scale, higher values indicate greater perceived fatigue). Error bars indicate 95\% confidence intervals.}
  \label{fig:graph_questionnaire}
\end{figure*}

\subsection{User Experience (Figure~\ref{fig:graph_questionnaire})}

\subsubsection{Task Load}
The analysis revealed a significant main effect of depth setting method ($F(1.54,35.46)=20.41,\ p<.001,\ \eta_p^2=.47$) on total workload. Ray Intersection yielded higher workload than Drag\&Hold ($t(23)=5.10$, $p<.001$, $d_z=1.04$) and Relative Gain ($t(23)=4.66$, $p<.001$, $d_z=.95$), whereas Drag\&Hold and Relative Gain did not differ ($p=1$).
The main effect of direction setting modality ($F(1,23)=0.11,\ p=.739,\ \eta_p^2=.00$) and the modality $\times$ method interaction ($F(1.91,43.82)=1.17,\ p=.317,\ \eta_p^2=.05$) were not significant.

Across six subscales, the depth setting method showed consistent effects. For Mental Demand, Effort, and Frustration, Ray Intersection yielded higher scores than both Drag\&Hold and Relative Gain ($ps \le .001$), whereas Drag\&Hold and Relative Gain did not differ ($ps \ge .405$). For Temporal Demand, Ray Intersection yielded a higher score than Relative Gain ($t(23)=2.74,\ p=.035,\ d_z=.56$), while the other method pairs were not significant ($ps \ge .054$). For Performance, Gaze yielded a worse performance than NDH ($F(1,23)=7.21,\ p=.013,\ \eta_p^2=.24$), and Ray Intersection yielded worse performance than both Drag\&Hold ($t(23)=2.79,\ p=.031,\ d_z=.57$) and Relative Gain ($t(23)=4.13,\ p=.001,\ d_z=.84$), whereas Drag\&Hold and Relative Gain did not differ ($p=.303$). For Physical Demand, depth setting method also showed a significant effect ($F(1.87,42.99)=9.19,\ p=.001,\ \eta_p^2=.29$). Ray Intersection yielded higher scores than Drag\&Hold ($t(23)=3.93,\ p=.002,\ d_z=.80$), and Relative Gain also yielded higher scores than Drag\&Hold ($t(23)=2.73,\ p=.036,\ d_z=.56$), whereas Ray Intersection and Relative Gain did not differ ($p=.228$).
No subscale showed a significant direction setting modality $\times$ depth setting method interaction ($ps \ge .169$).

\subsubsection{Usability}
The analysis revealed a significant main effect of depth setting method ($F(1.52,34.97)=28.43,\ p<.001,\ \eta_p^2=.55$). The score was lower for Ray Intersection than Drag\&Hold ($t(23)=5.34,\ p<.001,\ d_z=1.09$) and Relative Gain ($t(23)=6.41,\ p<.001,\ d_z=1.31$), whereas Drag\&Hold and Relative Gain did not differ ($p=1$). 
The main effect of direction setting modality ($F(1,23)=1.45,\ p=.241,\ \eta_p^2=.06$) and the direction setting modality $\times$ depth setting method interaction ($F(1.96,44.98)=2.52,\ p=.093,\ \eta_p^2=.10$) were not significant.

\subsubsection{Physical Fatigue}
\paragraph{Hand Fatigue}
The analysis revealed a significant main effect of direction setting modality ($F(1,23)=5.16,\ p=.033,\ \eta_p^2=.18$). Hand fatigue was higher for NDH than Gaze.
Depth setting method also showed a significant main effect ($F(1.83,42.08)=11.57,\ p<.001,\ \eta_p^2=.33$). Ray Intersection yielded higher hand fatigue than Drag\&Hold ($t(23)=4.63,\ p<.001,\ d_z=.94$), and Relative Gain yielded higher hand fatigue than Drag\&Hold ($t(23)=3.15,\ p=.013,\ d_z=.64$), whereas Ray Intersection and Relative Gain did not differ ($p=.190$).
%
The modality $\times$ method interaction was not significant ($F(1.87,42.99)=0.49,\ p=.603,\ \eta_p^2=.02$).

\paragraph{Eye Fatigue}
The analysis revealed a significant main effect of direction setting modality ($F(1,23)=14.41,\ p=.001,\ \eta_p^2=.39$). Eye fatigue was higher for Gaze than NDH.
%
Depth setting method also showed a significant main effect ($F(1.75,40.26)=6.83,\ p=.004,\ \eta_p^2=.23$). Ray Intersection yielded higher eye fatigue than Drag\&Hold ($t(23)=2.81,\ p=.030,\ d_z=.57$) and Relative Gain ($t(23)=3.03,\ p=.018,\ d_z=.62$), while Drag\&Hold and Relative Gain did not differ ($p=1$).
%
The modality $\times$ method interaction was not significant ($F(1.92,44.17)=0.77,\ p=.462,\ \eta_p^2=.03$).

\begin{figure}[tbp]
  \centering
  \includegraphics[width=0.925\linewidth]{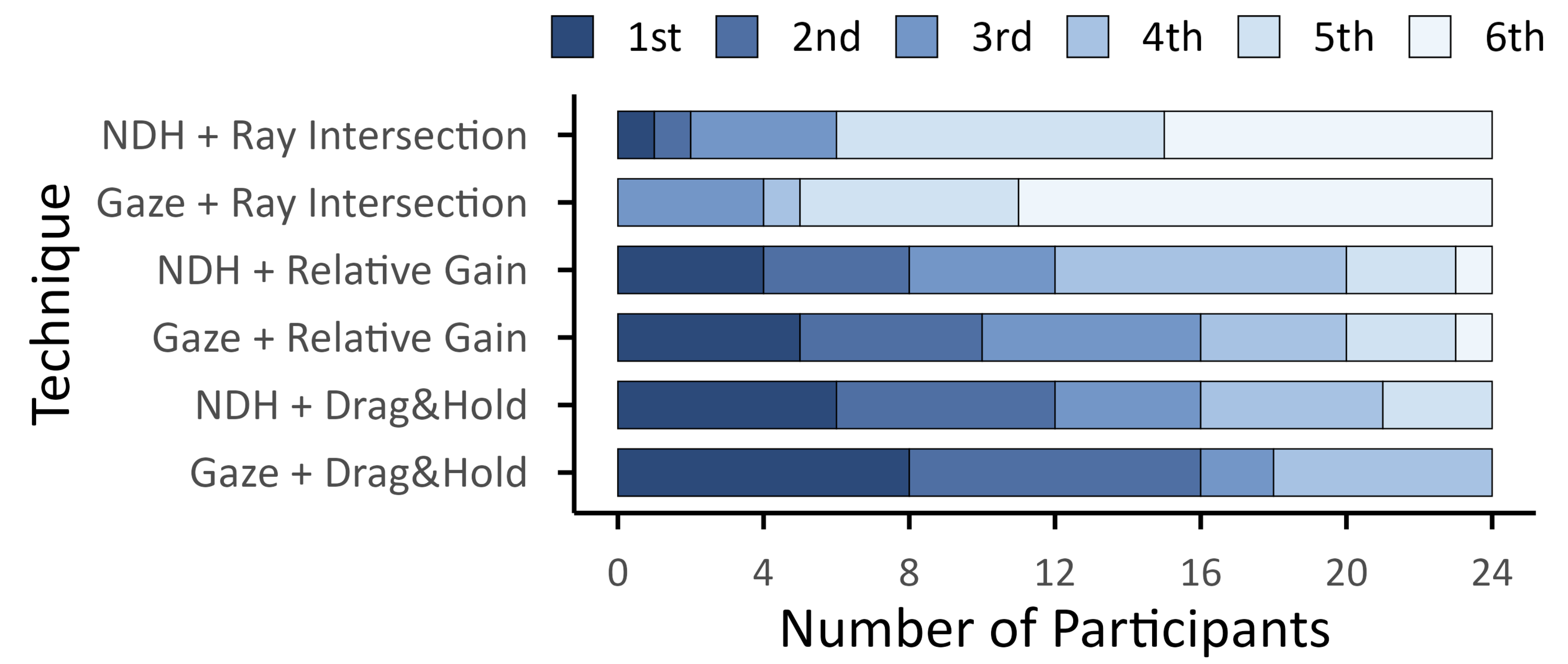}
  \caption{Preference ranking (1: most preferred, 6: least preferred).}
  \label{fig:graph_ranking}
\end{figure}

\subsection{Preference Ranking (Figure~\ref{fig:graph_ranking})}
Participants ranked the six techniques in order of preference. 
A Friedman test indicated a significant difference in preference rankings across the six techniques ($\chi^2(5)=47.69,\ p<.001$, Kendall's $W=.397$). Gaze+Drag\&Hold had the most favorable mean rank (2.25), followed by NDH+Drag\&Hold (2.71), Gaze+Relative Gain (2.92), and NDH+Relative Gain (3.21), whereas NDH+Ray Intersection (4.75) and Gaze+Ray Intersection (5.17) were least preferred. Post-hoc Nemenyi tests showed that both Ray-based techniques were ranked significantly lower than all four relative techniques ($ps \le .049$), whereas modality-matched pairs within the same depth setting method did not differ significantly ($ps \ge .958$).

\section{Discussion}
\subsection{RQ1: Direction Setting Method Shaped Speed, Coarse Accuracy, and Physical Fatigue}
Direction setting method influenced not only the direction setting stage itself but also the spawning efficiency, coarse accuracy, and physical fatigue. 
The shoulder-referenced NDH ray outperformed the viewpoint-based Gaze ray with shorter spawning times and smaller coarse offsets. Under the current lock-on-commit design, this may reflect that the explicit body-referenced cue provided by the NDH ray could make the intended direction easier to externalize and maintain than with the gaze cursor~\cite{argelaguet2009visual, bashar2025depth3dsketch}.

The effect of direction setting method was also evident in accuracy. NDH yielded smaller coarse offsets, indicating that the baseline established during direction setting was directly tied to initial spawn quality. This interpretation was also reflected in the interviews, in which eight participants reported greater confidence in direction setting with NDH. However, the difference disappeared in the final offset, suggesting that refinement partially compensated for the method-related differences in coarse spawn quality. 

Interestingly, these performance differences did not translate into differences in subjective usability or workload. Preference showed a similar dissociation in relative conditions, where the Gaze-based techniques received slightly better mean ranks than the NDH-based techniques. This may indicate that, in an object-spawning task in which approximate placement can be refined later, some users valued lower hand involvement and physical comfort more than precision. Consistent with this interpretation, 10 participants described the advantage of Gaze for its single-handed operation and lower physical demand, rather than accuracy. 
Overall, NDH favored faster and more stable coarse spawning, whereas Gaze remained competitive when lower hand involvement was prioritized.

\subsection{RQ2: Relative Depth Setting Methods Better Supported Reference-free Mid-air Spawning}

Depth setting method strongly shaped both performance and user experience in reference-free mid-air spawning. Relative Gain and Drag\&Hold outperformed Ray Intersection, yielding shorter depth setting, coarse and total spawn times, smaller offsets, lower workload, and more favorable usability and preference rankings. Both relative methods also reduced eye fatigue, whereas lower physical demand and hand fatigue were specific to Drag\&Hold. This suggests that, for reference-free spawning, it is more effective to allow depth to be adjusted independently after the directional baseline is established, rather than requiring continued precise alignment between the two rays. In the interviews, 10 participants described the intersection-based method as overly sensitive, difficult to predict, and uncomfortable for fine-tuning. 
 In Gaze+Ray Intersection, a few participants also reported difficulty judging depth along an invisible line or understanding which directional line they were adjusting. 
 Together, these responses suggest that Ray Intersection imposed a burden of precise alignment, with Gaze adding an extra cost due to the invisibility of the directional reference. Four participants also reported unintended position changes during the transition from semi-pinch to pinch, suggesting that ray intersection may be especially sensitive around instantiation. Even with a shoulder-referenced ray to reduce wrist-driven fluctuation, the sensitivity remained. If retained, future designs could stabilize or briefly freeze the intersection point around pinch onset~\cite{wolf2020understanding}. 

%

A different trade-off emerged between the two relative methods. Relative Gain produced the smallest coarse and final offsets and the shortest refinement time, with five participants valuing its understandable mapping and fine control. Drag\&Hold supported efficient depth traversal with the lowest hand fatigue and best mean preference rank; 10 participants described it as fast, intuitive, and comfortable. Thus, Relative Gain better supports accuracy-oriented tasks, whereas Drag\&Hold favors rapid, low-effort traversal despite requiring more refinement. Both were well accepted and preferred over Ray Intersection.

Drag\&Hold supported efficient depth setting with lower perceived physical demand and hand fatigue, but this advantage was partly offset by longer refinement than Relative Gain. This suggests that Drag\&Hold was effective for quickly reaching the approximate target depth, but required more subsequent correction. 
Relative Gain, by contrast, supported a closer initial placement, thereby reducing subsequent corrections and improving accuracy. Taken together, these findings suggest that Relative Gain may be better suited to tasks in which accurate single-object spawning is important. 
Nevertheless, both relative methods were well accepted. Both showed lower workload and higher usability than Ray Intersection, and there was no significant difference in preferences between them. Thus, both Relative Gain and Drag\&Hold remain practical design alternatives, with distinct strengths depending on whether accuracy or comfort is prioritized.

\subsection{RQ3: Far Spawn Depth Amplified Differences Across Direction and Depth Setting Methods}



Greater spawn depth broadly increased time and offsets, but not uniformly across techniques. Its temporal penalty was more pronounced with Gaze than NDH in direction setting, coarse spawning, and total spawning, whereas its accuracy penalty varied by depth method, with Ray Intersection showing the largest increase. Greater depth also increased refinement time only for Ray Intersection, indicating that its cost carried over from coarse spawning into refinement; four participants likewise described the method as less stable at far depth. However, hand movement did not increase uniformly with depth. It increased for Relative Gain, remained stable for Drag\&Hold, and decreased for Ray Intersection despite its greater accuracy loss. This suggests that physical movement depended on the control mapping rather than distance alone, and that lower movement did not necessarily indicate more robust control.


\subsection{Design Guidelines for Reference-free Mid-air Object Spawning}
Based on our findings, we derive design guidelines for reference-free mid-air position specification during object spawning. 
These implications are based on the single-object, target-based task and tested Gaze and NDH conditions under the lock-on-commit design.

\noindent\textbf{Use relative depth control as the default within a staged position specification pipeline.}
Relative Gain and Drag\&Hold outperformed Ray Intersection in spawning time, accuracy, workload, usability, and preference. Thus, when direction and depth are specified separately, relative depth control should generally be preferred over intersecting rays.

\noindent\textbf{Prefer Relative Gain when coarse spawn accuracy matters.}
Relative Gain produced the smallest coarse and final offsets and the shortest refinement time, making it suitable when accurate initial placement and reduced subsequent correction are important.

\noindent\textbf{Prefer Drag\&Hold when rapid depth traversal or lower perceived physical demand matters more than initial accuracy.}
Drag\&Hold supported efficient depth setting, produced the lowest physical-demand and hand-fatigue ratings, and maintained stable hand movement across the tested depths. However, it required more refinement than Relative Gain.

\noindent\textbf{Choose direction setting method based on workload distribution and expected spawn depth.}
NDH is preferable when faster spawning, coarse accuracy, and temporal robustness across depth are prioritized, whereas Gaze is preferable when reducing hand involvement is more important. This trade-off reflects greater hand movement and hand fatigue with NDH, versus greater eye fatigue and temporal sensitivity to spawn depth with Gaze. 

\section{Limitations and Future Work}

This study was conducted in a controlled environment using a single-object spawning task to isolate the core position specification performance. Accordingly, future work should extend the evaluation to more realistic scenarios involving clutter or occlusion, spawning without target markers, orientation and scale adjustment, and sequential multi-object placement.
It would also be valuable to support serial spawning more efficiently. For example, in relative methods, depth could remain at the previously set value, rather than returning to the default point, when the user releases a pinch while still in a semi-pinch state.
In addition, we did not include a comparison with an instantiation-and-reposition workflow. Since our goal was to isolate tradeoffs within reference-free spawning, such a comparison was beyond the scope of this study. Future work could examine how these approaches differ in realistic placement scenarios.
%
We locked direction once depth setting began to reduce instability and isolate the effects of direction and depth setting, although this may have reduced users’ sense of flexibility. Future work should evaluate continuously updated direction setting methods, as well as sensitivity to the starting depth, thresholds, gain, and filtering settings.
Because we tested only two spawn depths, future work should evaluate additional levels to characterize distance effects beyond the near–far contrast.
Finally, the observed NDH advantage may reflect the stabilizing role of a body-referenced anchor compared with our eye-origin gaze implementation. Future work should disentangle the input modality from anchoring strategies and examine performance across varying tracking qualities.


\section{Conclusion}
This work framed reference-free mid-air object spawning as a pre-instantiation position specification problem and evaluated direction and depth control within a staged interaction. The depth setting method most strongly shaped performance and user experience: Relative Gain and Drag\&Hold outperformed Ray Intersection, whose accuracy degraded most at greater depth. Direction setting involved a different trade-off: the shoulder-referenced NDH ray improved spawning speed and coarse accuracy, whereas the viewpoint-based Gaze ray reduced hand involvement. Position refinement attenuated the difference in coarse accuracy. These findings support prioritizing relative depth control and retaining pre-instantiation refinement within the staged interaction.



\acknowledgments{This work was supported by the National Research Foundation of Korea (NRF) grant funded by the Korea government (MSIT) (RS-2025-00521923).}

\bibliographystyle{abbrv-doi}

\bibliography{template}
\end{document}